# Active wetting of epithelial tissues: modeling considerations

Ivana Pajic-Lijakovic*, Milan Milivojevic

University of Belgrade, Faculty of Technology and Metallurgy, Department of Chemical Engineering, Karnegijeva 4, Belgrade 11000, Serbia

Correspondence to: Ivana Pajic-Lijakovic, iva@tmf.bg.ac.rs

**Abstract**

Morphogenesis, tissue regeneration and cancer invasion involve transitions in tissue morphology. These transitions, caused by collective cell migration (CCM), have been interpreted as active wetting/de-wetting transitions. This phenomenon is considered on model system such as wetting of cell aggregate on rigid substrate which includes cell aggregate movement and isotropic/anisotropic spreading of cell monolayer around the aggregate depending on the substrate rigidity and aggregate size. This model system accounts for the transition between 3D epithelial aggregate and 2D cell monolayer as a product of: (1) tissue surface tension, (2) surface tension of substrate matrix, (3) cell-matrix interfacial tension, (4) interfacial tension gradient, (5) viscoelasticity caused by CCM, and (6) viscoelasticity of substrate matrix. These physical parameters depend on the cell contractility and state of cell-cell and cell matrix adhesion contacts, as well as, the stretching/compression of cellular systems caused by CCM. Despite extensive research devoted to study cell wetting, we still do not understand interplay among these physical parameters which induces oscillatory trend of cell rearrangement. This review focuses on these physical parameters in governing the cell rearrangement in the context of epithelial aggregate wetting.de-wetting, and on the modelling approaches aimed at reproducing and understanding these biological systems. In this context, we do not only review previously-published bio-physics models for cell rearrangement caused by CCM, but also propose new extensions of those models in order to point out the interplay between cell-matrix interfacial tension and epithelial viscoelasticity and the role of the interfacial tension gradient in cell spreading.

**Key words**: collective cell migration; cell residual stress accumulation, tissue surface tension, Marangoni effect, viscoelasticity

**1.Introduction**

The CCM on substrate matrix plays not only a pivotal role in the cell rearrangement necessary for the establishment of proper tissue organization, shape, and size, but also underlie spreading of tumor cells (Douezan et al., 2011; Batlle and Wilkinson, 2012; Beaune et al., 2014;2018; Pérez-González et al., 2019; Pajic-Lijakovic and Milivojevic, 2020). The cell rearrangement itself depends on the dynamics at the cell-matrix biointerface. Batlle and Wilkinson (2012) pointed to three types of mechanism which influence the dynamics at the cell-matrix biointerface such as: (1) strength of cell-cell and cell-matrix adhesion contacts, (2) cell contractions, and (3) cell signaling which has a feedback on gene expression and the state of single-cells. Cell signaling influences the single cell state and cell rearrangement. The single cell state accounts for interplay among: the cell contractility, strength of cell-cell, and strength of cell matrix adhesion contacts (Blanchard et al., 2019; Barriga and Mayor, 2019). Strength of cell-cell adhesion contacts and cell contractility influence the tissue surface tension and viscoelasticity caused by CCM (Pajic-Lijakovic and Milivojevic, 2022a;2022d; Devanny et al., 2021). These parameters and the strength of cell-matrix adhesion contacts are involved into the cell-matrix interfacial tension and the interfacial tension gradient (Pajic-Lijakovic and Milivojevic, 2022d). Interplay between these physical parameters leads to the cell residual stress accumulation which can reduce cell movement (Trepat et al., 2009; Pajic-Lijakovic and Milivojevic, 2021). Cell tractions during CCM on substrate matrix influence the matrix viscoelasticity which can result in the matrix residual stress accumulation (Pajic-Lijakovic and Milivojevic, 2020). The accumulation of the residual stress within the substrate matrix induces the matrix stiffening which has a feedback on cell movement. This phenomenon is known as a durotaxis (Alert and Trepat, 2020). Interplay between surface characteristics of multicellular systems accompanied by its viscoelasticity, as well as, the viscoelasticity of substrate matrix influence the cell rearrangement. The main goal of this theoretical consideration is to point to this interplay between these physical parameters on the cell aggregate wetting/de-wetting based on the bio-physical model developed here.

The main characteristic of cell rearrangement caused by CCM, recognized within various model systems, is oscillatory change in: cell velocity, resulted strain, and the cell residual stress (Serra-Picamal et al., 2012; Notbohm et al., 2016; Pajic-Lijakovic and Milivojevic, 2020;2022a). This oscillatory trend of cell rearrangement has been known as the mechanical waves. Those mechanical waves generated within the multicellular systems represent a part of low Reynolds number turbulence (Pajic-Lijakovic and Milivojevic, 2020;2022c). Besides multicellular systems, the low Reynolds number turbulence appears during the rearrangement of various soft matter systems. Groisman and Steinberg (1998;2000) pointed to the so called "elastic turbulence" which appears during low Reynolds flow of polymer solution as a consequence of its viscoelasticity. The stiffening of stretched polymer chains during flow induces the generation of flow instabilities in this case. However, multicellular systems are much complex and capable of self-rearrangement so their behaviour is somewhat differenct. Alert at al. (2021) discussed the phenomenon by introducing the term "active turbulence" in order to describe oscillatory trend of CCM. Pajic-Lijakovic and Milivojevic (2020) discussed the role of viscoelasticity of multicellular systems in this case by formulating the viscoelastic force which represents the consequence of an inhomogeneous distribution of cell residual stress and matrix residual stress.

Mechanical waves, generated during CCM, are related to the effective, long-time inertial effects. Despite extensive research devoted to study the oscillatory trend of cell rearrangement within various model systems such as: (1) free expansion of cell monolayers (Serra-Picamal et al., 2012; Nnetu et al., 2012), (2) rearrangement of confluent cell monolayers (Notbohm et al., 2016), (3) cell aggregate rounding after uniaxial compression (Mombash et al., 2005; Pajic-Lijakovic and Milivojevic, 2017), (4) fusion of two cell aggregate (Pajic-Lijakovic and Milivojevic, 2022a), we still do not understand properly the cause of the effective inertia. The effective inertia represents a consequence of the interplay between surface properties of multicellular system accompanied by its viscoelasticity (Notbohm et al., 2016; Pajic-Lijakovic and Milivojevic, 2022a). Notbohm et al. (2016) considered the effective inertia as a result of coupling between cellular strain and cell active stress related to the myosin concentration. Serra-Picamal et al. (2012) and Deforet et al. (2014) formulated stochastic particle-based simulations of cell rearrangement. Serra-Picamal et al. (2012) balanced the active propulsion force with cell elastic force and cell-matrix friction force. Deforet at al. (2014) accounted for the force of inertia and balanced it with friction, intercellular adhesions, and active propulsion. However, in order to describe oscillatory trend of cell rearrangement during the cell aggregate wetting/de-wetting it is necessary to include interplay between the surface characteristics and viscoelasticity of multicellular systems.

The main goal of this review is to discuss the oscillatory trend of cell spreading during cell aggregate wetting/de-wetting on rigid substrate matrix [1,3,4]. This model system is very interesting since it includes the transition between 3D epithelial aggregate and 2D cell monolayer caused by physical interactions occurred at cell-matrix biointerface. Beaune et al. (2018) and Pérez-González et al. (2019) observed oscillatory change in: (1) the velocity of aggregate, (2) monolayer area, (3) cell normal residual stress, and (4) cell traction force per surface area, but have not account for the effective inertia into their modeling consideration. We expanded our model for describing the low Reynolds number turbulence proposed for 2D CCM (Pajic-Lijakovic and Milivojevic, 2020) by including the cell-matrix interfacial tension and the interfacial tension gradient and formulated a new bio-physical model which included cell force and mass balance at a mesoscopic level. Based on modeling consideration, we would like to extract the main physical parameters responsible for the oscillatory trend of cell aggregate wetting/de-wetting. Based on this modeling consideration, we also proposed two additional dimensionless numbers besides Reynolds number to characterize the corresponding low Reynolds turbulence. These are the effective Weissenberg number and Weber number. The aim of this theoretical analysis is to provoke biologists to provide additional experiments in order to deepen understand the dynamics at the biointerface which is an origin of various diseases.

## 2. Wetting/de-wetting of epithelial aggregates on solid substrates

The main aspects of cell aggregate wetting/de-wetting on rigid substrate are extracted based on experimental data from the literature in order to formulate the bio-physical model. The cell aggregate wetting/de-wetting has been considered depending on: (1) the cell aggregate size, (2) rigidity of the substrate matrix, (3) cell-cell cohesiveness (related to the level of E-cadherin expression), and (4) cell-matrix adhesiveness (Douezan et al., 2011; Beauene et al., 2018; Pérez-González et al., 2019). The cell rearrangement has been quantified by: (1) the velocity of aggregate, (2) monolayer area, (3) cell normal residual stress, (4) cell traction force per surface area. The schematics representations of the aggregate wetting/de-wetting on rigid substrate are shown in **Figure 1**.

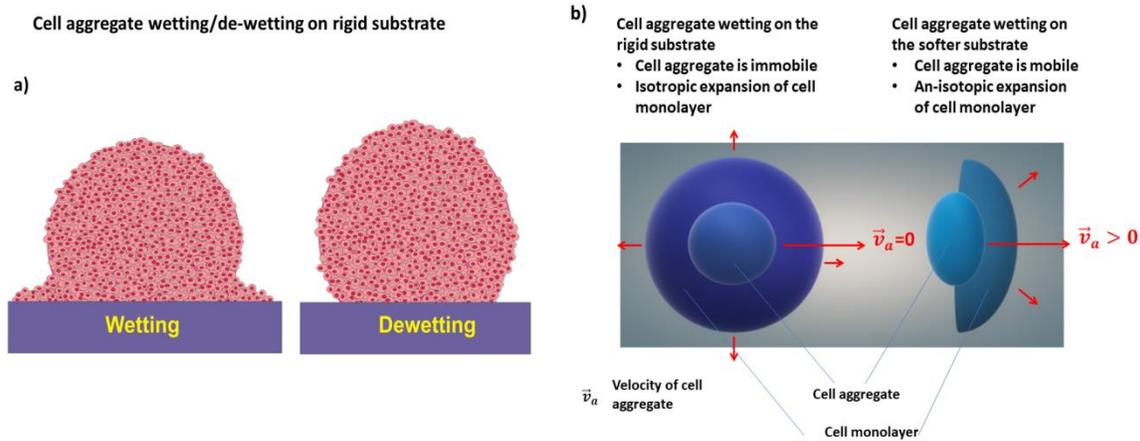

**Figure 1**. Behaviour of migrating cell aggregate on the substrate matrix: (a) cell aggregate wetting and (b) schematic presentation of cell spreading on the substrate depending on its rigidity.

Pérez-González et al. (2019) studied the rearrangement of confined and non-confined cell aggregates (which establish both E-cadherin mediated adherens junctions (AJs) and focal adhesions (FAs)) on collagen I gel and observed the transition from aggregate expansion (wetting) within $\sim 20\ h$ to their compression (de-wetting) within $\sim 40\ h$. Human breast adenocarcinoma cells (MDA-MB-231) stably transfected with a dexamethasone-inducible vector containing the human E-cadherin coding sequence were used. Pérez-González et al. (2019) pointed out that smaller cell aggregates de-wet earlier than larger ones [4]. It is in accordance with the fact that cell residual stress accumulation caused by CCM and the cell aggregate gravitation, pronounced for larger aggregates, can induce cell jamming at the aggregate-matrix contact area and on that base reduces cell spreading (Pajic-Lijakovic and Milivojevic, 2021;2022a). Consequently, only a part of cells (i.e. unjamming cells) actively contribute to the cell wetting.

The wetting is more intensive on rigid substrates since cells are able to establish FAs (Beaune et al., 2018). Beaune et al. (2018) considered rearrangement of murine sarcoma (E-cadherin) cell aggregates (with dieameter of ~100 μm) on fibronectin-coated polyacrylamide gels (PAA) with rigidities varying from 2 kPa to 40 kPa. While cell aggregate migrates on softer substrates, the aggregate is immobile on the rigid substrates (for the Young's modulus equal to $E = 40\ kPa$) (Beaune et al., 2018). In this case, the surrounding cell monolayer performs isotropic spreading away from the cell aggregate. While cell spreading is isotropic on rigid substrates, it becomes anisotropic on softer substrate (Beaune et al., 2018; Douezan and Brochard-Wyart, 2012). This interesting result could be related to the action of in-plane components of the aggregate gravitational force which can be significant on the softer substrate ($E = 16\ kPa$) (Beaune et al., 2018). The phenomenon will be discussed in the context of the bio-physical model formulated here. Further decrease in the substrate stiffness results in a decrease in the cohesiveness and stability of the cell monolayer around the aggregate. Cells are not able to establish strong-E cadherin mediated AJs on soft substrates ($E = 2\ kPa$) and perform single cell movement within the cell monolayer (Beaune et al., 2018). These cells undergo the epithelial-mesenchymal transition (Douezan et al., 2011). The cell monolayer consists of cohesive cells which establish AJs and FAs on rigid substrate and the substrate of the mediated rigidity (for Young's modulus of $10\ kPa < E < 20\ kPa$). The cell de-wetting occurs on rigid substrates rather than on softer ones (Beaune et al., 2018; Pérez-González et al., 2019). The main cause of such behaviour could be related to an increase in the epithelial surface tension caused by intensive stretching of cell monolayer on rigid substrates.

The bio-physical model is formulated in order to discuss interplay among physical parameters which influence cell spreading and extract the main parameters responsible for oscillatory trend of cell aggregate wetting/de-wetting.

### 3. Wetting/de-wetting of cell aggregates: the bio-physical model

Cell aggregate wetting/de-wetting is governed by interplay among physical parameters such as: (1) epithelial surface tension, (2) the surface tension of the substrate matrix, (3) cell-matrix interfacial tension, (4) the gradient of the interfacial tension, (5) cell residual stress accumulation, and (6) matrix residual stress accumulation. Interplay among these physical parameters is discussed based on the formulated bio-physical model which consists of three parts. Schematic presentation of the bio-physical model was shown in **Figure 2**.

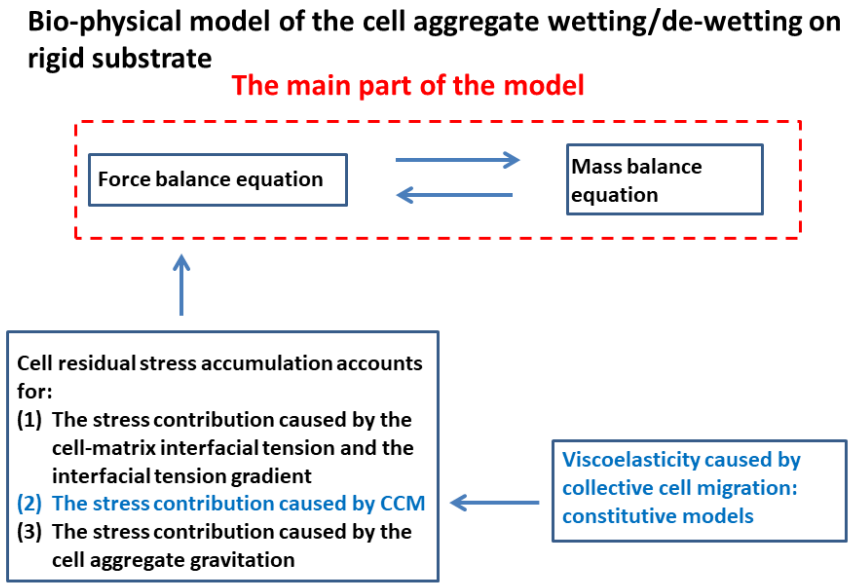

**Figure 2**. Schematic presentation of the bio-physical model for cell wetting/de-wetting on solid substrate.

The main part of the model includes cell force balance and mass balance. The second part of the model is related to the formulation of the cell residual stress which influences the viscoelastic force. The cell residual stress includes several contributions: (1) the stress contribution resulted by cell-matrix interactions at the biointerface quantified by the cell-matrix interfacial tension and interfacial tension gradient, (2) the stress contribution resulted by the cell aggregate gravitation which induces the compression of cells within the aggregate-matrix contact area, and (3) the stress contribution caused by CCM. The third part of the model accounts for the discussion of the cell residual stress accumulation caused by CCM in the context of various viscoelastic constitutive models proposed in the literature depending on the cell velocity and cell packing density. We will provide the short description of physical parameters such as: (1) epithelial surface tension, (2) matrix surface tension, (3) cell-matrix interfacial tension, (4) cell residual stress accumulation, and (5) matrix residual stress accumulation, and then include them into bio-physical model formulated here.

### 3.1 The surface tension of multicellular systems

The macroscopic tissue surface tension depends on the single-cell state and cell rearrangement caused by CCM. The relevant variables for describing the single cell state are: the strength of cell-cell adhesion contacts and cell contractility (Stirbat et al., 2013). Devanny et al. (2021) considered rearrangement of various breast cells and pointed out that epithelial cells, which establish E-cadherin mediated AJs, have much higher surface tension in comparison to the mesenchymal cells which establish weak AJs. This is important finding since one of the main physical factors which influence malignant behavior of tumor is its cohesiveness (Douezan et al., 2011). The strength of AJs among epithelial cells is enhanced by cell contractions which lead to an increase in the surface tension in comparison to non-contractile epithelial cells (Devanny et al., 2021).

Besides single cell state, the rearrangement of epithelial cells also influences the epithelial surface tension. Stretching of AJs caused by cell expansion (i.e. cell wetting) induces a reinforcement of E-cadherin mediated AJs which could lead to an increase in the surface tension. Otherwise, compression of epithelium (i.e. cell de-wetting) can intensify the contact inhibition of locomotion (CIL) accompanied by weakening of cell-cell adhesion contacts which leads to a decrease in the epithelial surface tension (Pajic-Lijakovic and Milivojevic 2022b).

The macroscopic tissue surface tension has been measured by cell aggregate uni-axial compression between parallel plates (Mombash et al., 2005; Marmottant et al., 2009). Two types of the surface tension can be distinguished: static and dynamic. While static surface tension describes the aggregate surface characteristics which correspond to the equilibrium state, reached out after uni-axial compression, the dynamics surface tension depends on the cell rearrangement pathway during the aggregate relaxation toward the equilibrium state. Both surface tensions can be quantified by measuring the aggregate geometry based on the Young-Laplace equation (Marmottant et al., 2009). The dynamic tissue surface tension depends on the change in the cell aggregate surface $\frac{\Delta A_a}{A_{a0}}$ during the relaxation process (where $A_{a0}$ is the initial surface of cell aggregate, $\Delta A_a$ is the surface change caused by cell rearrangement, and $\tau$ is the long-time scale, i.e. the time scale of hours which corresponds to CCM). The free expansion of cell monolayers and cell aggregate wetting on the rigid substrate also induce an increase in the multicellular surface which has a feedback on the tissue surface tension. Extensive studies have been devoted to measure and model the static surface tension of various multicellular systems (Mombash et al., 2005; Stirbat et al., 2013). However, much less attention has been paid to consider the dynamic tissue surface tension based on the dilational viscoelasticity (Babak et al., 2005). The inter-relation between the dynamic tissue surface tension and the change in the multicellular system caused by CCM can be formulated based on several characteristics of the epithelial surface rearrangement:

- The epithelial surface tension $\gamma_c$ causes a decrease in the surface of cell aggregate $\frac{\Delta A_a}{A_{a0}}$ toward the equilibrium state. The phenomenon has been considered on the model system such as the fusion of two cell aggregates (Pajic-Lijakovic and Milivojevic, 2022a;2022e). In this case, the surface of two-aggregate systems relaxes exponentially (Pajic-Lijakovic and Milivojevic, 2022e). The cell aggregate shape and corresponding surface also relax exponentially after the aggregate uni-axial compression (Mombash et al., 2005). This exponential relaxation points to a liner constitutive model as appropriate for modeling those systems. The surface relaxation of multicellular systems occurs via CCM and corresponds to a time scale of hours (Marmottant et al., 2009).
- The free expansion of epithelial monolayer during cell aggregate wetting on rigid substrate induces stretching of AJs which leads to an increase in the epithelial surface tension (Devanny et al., 2021). In contrast to expansion, the cell monolayer compression (i.e. de-wetting) leads to an intensive CIL which results in a decrease in the epithelial surface tension (Pajic-Lijakovic and Milivojevic, 2022b). Consequently, the magnitude of the surface tension is directly proportional to the surface deformation of the multicellular system.

Based on these findings (related to the relationship between $\gamma_c$ and $\frac{\Delta A_a}{A_{a0}}$ from various cellular model systems) the corresponding linear constitutive model for the local cell surface rearrangement caused by CCM can be expressed in the form:

$$\Delta \gamma_c = E_S \frac{\Delta A_a}{A_{a0}} + \eta_S \frac{d}{d\tau}\left(\frac{\Delta A_a}{A_{a0}}\right) \qquad (1)$$

where $\Delta \gamma_c = \gamma_c - \gamma_{c0}$ and $\gamma_{c0}$ is the initial surface tension, $E_S$ is the surface modulus of elasticity, and $\eta_S$ is the surface viscosity. The epithelial surface tension accompanied by the viscoelasticity caused by CCM is the one of key parameters which influences the cell aggregate wetting/de-wetting on rigid substrate.

It is necessary to point out the order of magnitude of the tissue surface tension and compare it with the surface tension of substrate matrix because both surface tensions accompanied by the cell-matrix interfacial tension influence cell wetting/de-wetting.

### 3.1.1 The tissue surface tension and matrix surface tension: the order of magnitude

Mombach et al. (2005) considered a change in tissue surface tension during the rounding of 3D chicken embryonic neural retina spheroids. Corresponding tissue surface tension increases from 1.6±0.6 mN/m to 4.0±1.0 mN/m within 9 days. Beysens et al. (2000) found that the values of the average tissue surface tension for five chicken embryonic tissues varied from 1.6 mN/m (neural retina) to 20 mN/m (limb bud). The surface tension of F9 WT cell aggregate is 4.74±0.28 mN/m (Stirbat et al., 2013).

Collagen I gel, among others, has been frequently used as a substrate matrix. Surface tension of collagen I gel depends on the surface distribution of hydrophobic parts of the filaments. Baier and Zisman (1975) reported that the surface tension of native rat skin collagen is 40 mN/m. Surface tension of collagen I gel decreases with increasing the collagen concentration (Kezwon and Wojciechowski, 2014). Corresponding increase in the collagen concentration from $10^{-7} \frac{mol}{dm^3}$ to $10^{-5} \frac{mol}{dm^3}$ induces a decrease in the surface tension from 72 mN/m to 58 mN/m. Consequently, cell tractions as well as their movement toward the substrate induce the substrate compression which have a feedback on the surface tension of collagen I hydrogel, as well as, on the interfacial tension between the epithelial aggregate and the matrix. The surface tension of collagen I gel $\gamma_M$ satisfies the condition $\gamma_M > \gamma_c$. In further consideration it is necessary to discuss the cell and matrix residual stress accumulation caused by cell rearrangement on the substrate matrix as the one of the main contributors to the oscillatory dynamics of cell aggregate wetting/de-wetting.

### 3.2 Cell-matrix interfacial tension

While the epithelial and matrix surface tensions have been measured, the cell-matrix interfacial tension has not been measured yet. The epithelial surface tension $\gamma_c$ primarily depends on the state of AJs, while the cell-matrix interfacial tension $\gamma_{cM}$ depends on the strength of FAs. Consequently, the state of FAs (and the interfacial tension $\gamma_{cM}$) depends on the rigidity of the substrate matrix. The FAs and AJs accompanied by the actin cytoskeleton represent an inter-connected system responsible for the transfer of forces between cell and its surroundings (Zuidema et al., 2020). An increase in the strength of FAs results in an increase in tension on AJs (Devanny et al., 2021; Wang et al., 2006). The inter-relation between epithelial surface tension $\gamma_c$ and the cell-matrix interfacial tension $\gamma_{cM}$ for the cell aggregate wetting and de-wetting will be discussed in the context of the formulated bio-physical model.

### 3.3 The residual stress accumulation within multicellular system and the substrate matrix

The cell residual stress is generated as consequence of: (1) cell-matrix interactions at the biointerface quantified by the interfacial tension and interfacial tension gradient, (2) cell rearrangement caused by CCM, and (3) cell aggregate gravitation. The residual stresses within the cell aggregate and the substrate matrix include: normal stress and shear stress contributions. The cell normal residual stress $\tilde{\sigma}_{crV}$ accounts for isotropic and deviatoric parts and is formulated based on modified model proposed by Pajic-Lijakovic and Milivojevic (2022c):

$$\tilde{\sigma}_{crV} = \pm \Delta p_{c \to M} \tilde{I} + \tilde{\sigma}_{crV}{}^d \tag{2}$$

where $\Delta p_{c \to M}$ is the isotropic part, $\tilde{I}$ is the unit tensor, and $\tilde{\sigma}_{crV}{}^d$ is the deviatoric part of the cell normal stress at the cell-matrix biointerface. De-wetting (compression) is labeled by sign "+", while the wetting (extension) is labeled by sign "-". The isotropic part of the cell normal residual stress is generated by the work of the interfacial tension $\gamma_{cM}$ and was expressed based on the Young-Laplace equation in the form (Pajic-Lijakovic and Milivojevic, 2022c):

$$\Delta p_{c \to M} = -\gamma_{cM}(\vec{\nabla} \cdot \vec{n}) \tag{3}$$

where $\vec{n}$ is the normal vector of the interface. The deviatoric part of cell normal residual stress $\tilde{\sigma}_{crV}{}^d$ includes two contributions: (1) the stress caused by CCM $\tilde{\sigma}_{crV}{}^{CCM}$ and (2) the stress generated by the z-component of the aggregate gravitational force $\tilde{\sigma}_G{}^d$:

$$\tilde{\sigma}_{crV}{}^d = \tilde{\sigma}_{crV}{}^{CCM} + \tilde{\sigma}_G{}^d \tag{4}$$

When cells undergo movement toward the contact area between the aggregate and substrate (i.e. wetting), cells compressed the matrix. Otherwise, when cells undergo de-wetting this matrix compression is reduced. Consequently, the matrix normal residual stress accounts for isotropic and deviatoric parts:

$$\tilde{\sigma}_{MrV} = \mp \Delta p_{c \to M} \tilde{I} + \tilde{\sigma}_{MrV}{}^d \tag{5}$$

where $\tilde{\sigma}_{MrV}{}^d$ is the matrix deviatoric stress which includes two contributions: (1) the stress contribution caused by cell traction $\tilde{\sigma}_{MrV}{}^{TR}$ and (2) stress contribution caused by action of the z-component of the aggregate gravitational force $\tilde{\sigma}_{MV}{}^G$. Consequently, the deviatoric part of the matrix stress can be expressed as: $\tilde{\sigma}_{MrV}{}^d = \tilde{\sigma}_{MrV}{}^{TR} + \tilde{\sigma}_{MV}{}^G$. The matrix residual stress accumulation induces stiffening of the substrate matrix depended on the matrix viscoelasticity which has a feedback on cell movement.

The cell shear stress generated at the contact area between the aggregate and substrate matrix accounts for two contributions, i.e. the cell shear stress generated by natural and forced convections. The natural convection is guided by the corresponding gradient of the interfacial tension expressed as: $\vec{\nabla}\gamma_{cM} = \frac{\gamma_{Mc} - \gamma_c}{\Delta_c}\vec{t}$ (where $\Delta_c$ is the thickness of the perturbed layer of the epithelial cells within the contact area and $\vec{t}$ is the tangent vector of the interface). It is a part of the Marangoni effects related to a movement of system constituents from the regions of lower surface tension to the regions of larger surface tension by natural convection (Pajic-Lijakovic and Milivojevic, 2022d). The Marangoni effect exists in various soft matter systems. The surface tension gradient can be induced by changing the temperature or distribution of constituents within the soft matter system (Karbalaei et al., 2016). The forced convection includes two cell stress contributions: (1) the cell shear residual stress caused by CCM and (2) the cell shear stress caused by in-plane action of the aggregate gravitational force. The shear stress generated at the cell aggregate-matrix contact area is expressed by modified model proposed by Pajic-Lijakovic and Milivojevic (202022c;2022d):

$$\vec{n} \cdot \tilde{\sigma}_{crS} \cdot \vec{t} = \vec{\nabla}\gamma_{cM} \cdot \vec{t} + \vec{n} \cdot \tilde{\sigma}_{crS}^{CCM} \cdot \vec{t} + \vec{n} \cdot \tilde{\sigma}_{cS}^{G} \cdot \vec{t} \qquad (6)$$

where $\tilde{\sigma}_{crS}^{CCM}$ is the cell shear stress generated by CCM.

The shear stress accumulation within the matrix is caused by: (1) cell traction and (2) the matrix deformation caused by the aggregate gravitational force. Therefore the matrix shear stress generated is equal to:

$$\tilde{\sigma}_{MrS} = \tilde{\sigma}_{MrS}^{TR} + \tilde{\sigma}_{MS}^{G} \; . \qquad (7)$$

where $\tilde{\sigma}_{MrS}^{TR}$ is the matrix shear stress induced by cell tractions which depends on the matrix viscoelasticity and $\tilde{\sigma}_{MS}^{G}$ is the additional component of the matrix shear stress caused by the in-plane components of the aggregate gravitational force.

**3.4 Cell residual stress accumulation caused by CCM**

Cell residual stress accumulation caused by CCM is a product of viscoelastic nature of multicellular systems [37]. The viscoelasticity has been considered on two time-scales, i.e. short-time scale (the time scale of minutes) and long-time scale (the time scale of hours) (Pajic-Lijakovic and Milivojevic, 2019a;2020). Long time scale corresponds to CCM, resulted generation of strain and its change, as well as, the cell residual stress accumulation (Pajic-Lijakovic and Milivojevic, 2019a;2020). Short time scale corresponds to the cell stress relaxation time [28]. Consequently, cell stress relaxes during successive short time relaxation cycles, while strain change caused by CCM occurs at a long time scale (Pajic-Lijakovic and Milivojevic, 2020).

Generated stress within multicellular systems caused by CCM can be normal and shear (Tambe et al., 2013; Beauene et al., 2014; Pajic-Lijakovic and Milivojevic, 2022d). Normal stress is accumulated within migrating epithelial clusters and during the collision among clusters caused by uncorrelated motility (Pajic-Lijakovic and Milivojevic, 2019a;2020). The uncorrelated motility is pronounced for the case of larger cell aggregates which causes more intensive cell residual stress accumulation during the cell aggregate wetting on rigid substrate (Pérez-González et al., 2019). The cell spreading on the rigid substrate is more intensive and consequently leads to a generation of higher cell normal stress in comparison to the cell spreading on softer substrate (Pérez-González et al., 2019). Shear stress can be significant within the biointerface between migrating cell clusters and surrounding epithelial cells in the resting state within the aggregate core region or substrate matrix (Pajic-Lijakovic and Milivojevic, 2019a;2022d). Cell residual stress accumulated during CCM can be elastic or dissipative depending on the interplay between single-cell state and cell rearrangement. The single-cell state is related to the state of cell-cell adhesion contacts and cell contractility. The cell rearrangement depends on cell packing density and cell velocity and has a feedback on single-cell state (Trepat et al., 2009; Pajic-Lijakovic and Milivojevic, 2021). In order to express the cell residual stress, it is necessary to establish proper constitutive, viscoelastic model.

### 3.4.1 Constitutive, viscoelastic models for cell rearrangement caused by CCM

Some authors have treated migrating epithelial collectives as viscoelastic liquids (Beauene et al., 2014; Pérez-González et al., 2019;Oswald et al., 2017). This assumption has been supported based on the fact that epithelial systems have the tissue surface tension and that the surface tension represents the characteristic of liquids. However, amorphous solids such as polymer hydrogels and foams also have a surface tension (Mondal et al., 2015). The viscoelasticity of epithelial systems caused by CCM depends on cell contractility and the state of AJs. In order to characterize the viscoelasticity by proper constitutive models, it is necessary to take into consideration experimental data such as cell velocity, corresponding strain, cell packing density, stress change caused by CCM obtained for various model systems in the literature.

Guevorkian et al. (2011) considered epithelial aggregate micropipette aspiration. Forced cell movement into the pipette induces disruption of cell-cell AJs accompanied by intensive energy dissipation which is characteristic of the viscoelastic liquids and imply use of the Maxwell model. Corresponding cell velocity is $\vec{v}_c > 1 \frac{\mu m}{min}$. However, the rearrangement of various epithelial multicellular systems caused by CCM shows viscoelastic solid behaviours such as: (1) the free expansion of epithelial monolayers (Serra-Picamal et al., 2012), (2) rearrangement of confluent epithelial monolayers (Notbohm et al., 2016), and (3) cell aggregate rounding after uni-axial compression between parallel plates (Mombash et al., 2005; Marmottant et al., 2009). Marmottant et al. (2009) measured stress relaxation under constant strain conditions and strain relaxation under constant stress (or zero stress) conditions during cell aggregate uni-axial compression between parallel plates. The stress and strain for this system relax exponentially which points to a liner constitutive model. Stress relaxation time corresponds to a time scale of minutes, while the strain relaxation time corresponds to a time scale of hours (Marmottant et al., 2009). The strain relaxation under constant stress condition represents a confirmation of viscoelastic solid behaviour (Pajic-Lijakovic and Milivojevic, 2021). Serra-Picamal et al. (2012) and Notbohm et al. (2016) revealed that cell residual stress correlates with the corresponding strain caused by CCM. This characteristic pointed to the Zener model as suitable for describing the cell rearrangement caused by CCM. Ability of strain to relax under constant stress conditions and stress to relax under constant strain conditions also pointed to the Zener constitutive model.

Maximum cell velocity obtained during free expansion of the Madin-Darby canine kidney (MDCK) type II cell monolayer is $\vec{v}_c = 0.5 \frac{\mu m}{min}$ (Serra-Picamal et al., 2012), while the maximum cell velocity obtained during the rearrangement of confluent MDCK monolayers is $\vec{v}_c = 0.25 \frac{\mu m}{min}$ (Notbohm et al., 2016). Beauene et al. (2018) considered the wetting of cell aggregate on rigid substrate. The cell aggregate can be mobile or immobile, while the cell monolayer performs isotropic or anisotropic spreading around the aggregate depending on the substrate rigidity. The average velocity of cells within the monolayer was $\vec{v}_c = 0.48 \frac{\mu m}{min}$, which is similar to the maximum velocity obtained during free expansion of MDCK cell monolayers (Serra-Picamal et al., 2012). It was expectable that the spreading of cohesive cell monolayer around the cell aggregate on rigid substrate corresponds to free expansion of cell monolayers. However, collision of cell velocity fronts caused by uncorrelated motility can induce local increase in the cell normal residual stress accumulation accompanied by an increase in cell packing density which can lead to the cell jamming state transition (Pajic-Lijakovic and Milivojevic, 2019a). The main characteristics of migrating cell collectives are in-homogeneous distributions of the cell packing density, velocity, and cell residual stress (Serra-Picamal et al., 2012; Nnetu et al., 2012; Notbohm et al., 2016). Consequently, some parts of the epithelial systems are in the jamming state, while the other part actively migrates (Nnetu et al., 2012; Pajic-Lijakovic and Milivojevic, 2021). Cells are jammed some period of time and then start migrating again.

Pajic-Lijakovic and Milivojevic (2021) established several viscoelastic regimes for the epithelial multicellular systems such as: (1) convective, (2) conductive, and (3) damped-conductive regimes depending on the cell velocity and packing density. The convective regime accounts for two sub-regimes. The convective sub-regime 1 corresponds to a viscoelastic liquid behaviour, described by the Maxwell model, for cell velocity $\vec{v}_c > 1 \frac{\mu m}{min}$ and cell packing density $n_c < n_{conf}$ (where $n_{conf}$ is the cell packing density at confluent state). The convective sub-regime 2 corresponds to a viscoelastic solid behaviour, described by the Zener model, for lower cell velocity, i.e. $0.1 \frac{\mu m}{min} < \vec{v}_c < 1 \frac{\mu m}{min}$ and the same cell packing density $n_c \leq n_{conf}$. The viscoelastic regimes characterized by cell velocity and packing density accompanied by the proper constitutive models for their description are shown in **Table 1**. The cell normal and shear residual stresses $\tilde{\sigma}_{crV}{}^{CCM}$ and $\tilde{\sigma}_{crS}{}^{CCM}$ represent the parts of eqs. 4 and 6, respectively.

**Table 1**. Constitutive models for various viscoelastic regimes

| Epithelial viscoelastic regime | Cell velocity and packing density of epithelial cells | Constitutive model for the rearrangement of epithelial cells |
|---|---|---|
| Convective regime | $\vec{v}_c > 1 \frac{\mu m}{min}$ <br> $n_c \leq n_{con}$ <br> ($n_{con}$ is the cell packing density at confluent state) | The Maxwell model <br> $\tilde{\sigma}_{ci}(\Re, t_s, \tau)^{CCM} + \tau_{Ri}\dot{\tilde{\sigma}}_{ci}(\Re, t_s, \tau) = \eta_{ci}\dot{\tilde{\varepsilon}}_{ci}(\Re, \tau)$ <br> Cell residual stress is dissipative. <br> $\tilde{\sigma}_{cri}(\Re, \tau)^{CCM} = \eta_{ci}\dot{\tilde{\varepsilon}}_{ci}$ |
| | $0.1 \frac{\mu m}{min} < \vec{v}_c < \sim 1 \frac{\mu m}{min}$ <br> $n_c \leq n_{con}$ | The Zener model <br> $\tilde{\sigma}_{ci}(\Re, t_s, \tau)^{CCM} + \tau_{Ri}\dot{\tilde{\sigma}}_{ci}(\Re, t_s, \tau) = E_{ci}\tilde{\varepsilon}_{ci}(\Re, \tau) + \eta_{ci}\dot{\tilde{\varepsilon}}_{ci}(\Re, \tau)$ <br> Cell residual stress is elastic. <br> $\tilde{\sigma}_{cri}(\Re, \tau)^{CCM} = E_{ci}\tilde{\varepsilon}_{ci}$ |
| Convective regime | $10^{-3} \frac{\mu m}{min} < \vec{v}_c < 10^{-2} \frac{\mu m}{min}$ <br> $n_j > n_c > n_{con}$ <br> ($n_j$ is the cell packing density at jamming state) | The Kelvin-Voigt model <br> $\tilde{\sigma}_{ci}(\Re, \tau)^{CCM} = E_{ci}\tilde{\varepsilon}_{ci} + \eta_{ci}\dot{\tilde{\varepsilon}}_{ci}$ <br> The stress cannot relax. <br> $\tilde{\sigma}_{ci}(\Re, \tau)^{CCM} = \tilde{\sigma}_{cri}$ |

| Damped-conductive regime | $\vec{v}_e \to 0$<br>$n_c \to n_j$ | The Fraction model<br>$\tilde{\sigma}_{ci}(\Re,\tau)^{CCM} = \eta_{\alpha i} D^{\alpha}(\tilde{\varepsilon}_{ci})$, $\alpha \leq 0.5$<br>The stress cannot relax.<br>$\tilde{\sigma}_{ci}(\Re,\tau)^{CCM} = \tilde{\sigma}_{cri}$ |
|---|---|---|

where $i \equiv S, V$, $S$ is shear, $V$ is volumetric, $t_s$ is the short time scale (i.e. a time scale of minutes), $\Re = \Re(x,y,z)$ represents the coordinate of cells within the aggregate-substrate contact area and within the monolayer, $\tilde{\sigma}_{cri}(\Re,\tau)^{CCM}$ is the cell residual stress accumulation caused by CCM, $\vec{u}_c$ is the cell displacement field, $\tilde{\sigma}_{ci}$ is the cell stress (shear or normal), $\tilde{\varepsilon}_{ci}$ is the strain (shear or volumetric), $\tilde{\varepsilon}_{cS} = \frac{1}{2}\left(\vec{\nabla}\vec{u}_c + \vec{\nabla}\vec{u}_c^T\right)$ is the shear strain, $\tilde{\varepsilon}_{cV} = (\vec{\nabla} \cdot \vec{u}_c)\tilde{I}$ is the volumetric strain, $\tilde{I}$ is the unity tensor, $\dot{\tilde{\varepsilon}}_{ci}$ is the strain rate, $\dot{\tilde{\sigma}}_{ci}$ is the rate of stress change, $E_{ci}$ is the Young's or shear modulus, $\eta_{ci}$ is shear or bulk viscosity, $D^{\alpha}\tilde{\varepsilon}(\Re,\tau) = \frac{d^{\alpha}\tilde{\varepsilon}(\Re,\tau)}{d\tau^{\alpha}}$ is the fractional derivative, and $\alpha$ is the orders of fractional derivative (the damping coefficient), $\eta_{\alpha i}$ is the effective modulus (volumetric or shear) for the transient and jamming sub-regimes. Caputo's definition of the fractional derivative of a function $\tilde{\varepsilon}(\Re,\tau)$ was used, and it is given as: $D^{\alpha}\tilde{\varepsilon} = \frac{1}{\Gamma(1-\alpha)}\frac{d}{dt}\int_0^t \frac{\tilde{\varepsilon}(\Re,\tau')}{(\tau-\tau')^{\alpha}}d\tau'$ (where $\Gamma(1-\alpha)$ is a gamma function) (Podlubny, 1999).

Corresponding cell residual stress for the Zener model is purely elastic (**Table 1**). The maximum cell residual stress (normal and shear) accumulated during free expansion of MDCK cell monolayers corresponds to $100 - 150\ Pa$ (Tambe et al., 2013), while the maximum cell normal residual stress caused by the rearrangement of confluent monolayers of MDCK cells is $\sim 300\ Pa$ (Notbohm et al., 2016). Accumulation of cell normal residual stress leads to an increase in cell packing density and decrease in cell mobility which results in change of the viscoelasticity from convective to conductive regime (Trepat et al., 2009; Pajic-Lijakovic and Milivojevic, 2021).

Conductive regime corresponds to a higher cell packing density $n_j > n_c > n_{conf}$ and lower (diffusion) cell velocity $\vec{v}_c \leq 10^{-2}\frac{\mu m}{min}$ (where $n_j$ is the cell packing density under the jamming state). In this regime cell movement occurs via linear, diffusion mechanism and viscoelasticity corresponds to viscoelastic solids described by the Kelvin-Voigt model (Pajic-Lijakovic and Milivojevic, 2021;2022b) (**Table 1**). The main characteristic of this regime is that stress cannot relax. Corresponding long-time change in cell stress accounts for elastic and dissipative parts. Angelini et al. (2011) considered movement of MDCK cells and pointed out that cell diffusion coefficient decreases from $\sim 0.40\ \frac{\mu m^2}{min}$ to $\sim 0.10\ \frac{\mu m^2}{min}$ when the packing density of MDCK cells increases from $\sim 1.40 x 10^5\ \frac{cells}{cm^2}$ to $\sim 2.63 x 10^5\ \frac{cells}{cm^2}$. The diffusion coefficient of collectively migrating endodermal cells is equal to $D_{eff} = 1.05 \pm 0.4\ \frac{\mu m^2}{min}$ (Rieu et al., 2000). Beaune et al. (2018) considered cell aggregate wetting and characterized the movement of cell aggregate as a conductive with the diffusion coefficient expressed as $D_{eff} = v_a R_a$ (where $v_a$ is the average velocity of the aggregate center of mass and $R_a$ is the aggregate radius). For the average velocity $v_a \sim 0.48\ \frac{\mu m}{min}$ and the aggregate radius equal to $R_a = 70\ \mu m$, the corresponding diffusion coefficient is equal to $D_{eff} = 33.6\ \frac{\mu m^2}{min}$. This is very large value. Douezan and Brochard-Wyart (2012) reported that the corresponding diffusion coefficient of movement of murine sarcoma (S-180) cells is $\sim 9\ \frac{\mu m^2}{min}$. Consequently, the cell wetting considered by Beaune et al. (2018) could rather correspond to the convective regime similarly as free expansion of cell monolayers, while conductive regime accompanied by the cell jamming can exist locally during cell rearrangement as a consequence of an increase of cell packing density caused by cell normal residual stress accumulation (Trepat et al., 2009; Pajic-Lijakovic and Milivojevic, 2021).

Further increase in cell packing density, for $n_c \to n_j$ and cell velocity $\vec{v}_c \to 0$, leads to a transition from the conductive to damped-conductive regime. This regime describes cell movement near the jamming state which occurs via non-linear, sub-diffusion mechanism (Nnetu et al., 2013; Pajic-Lijakovic and Milivojevic, 2022b). In this regime, cell collective behaves as a viscoelastic solid, that can be described by a non-linear, fractional constitutive model proposed by Pajic-Lijakovic and MIlivojevic (2019b;2021) (**Table 1**). Beaune et al. (2014) considered wetting/de-wetting of cell aggregates on collagen I substrate depending on the aggregate size and revealed that part of cells is in the jamming state. The part of cells in the jamming state during the aggregate wetting depends on the aggregate size. Only a part of cells within the surface region of larger aggregates (with the radius $R_0 > 65\ \mu m$) is capable of spreading, while the other part within the aggregate core region is in the jamming state. While only a part of cells is active within larger aggregates, almost all cells are active within smaller aggregates (Beaune et al., 2014).

**3.5 The force balance: modeling consideration**

The cell spreading within the film (i.e. wetting) is driven by the force of gravity, mixing force, and interfacial tension force, against the viscoelastic force, and traction force. The force of gravity is equal to $\vec{F}_{ga} = \rho_a \vec{g}$ (where $\rho_a$ is the aggregate density and $\vec{g}$ is the gravitational acceleration). The prerequisite of isotropic cell spreading within the film is that $\vec{g} = g_z \vec{k}$ (where $g_z$ is the z-component of the vector $\vec{g}$ and $\vec{k}$ is the unit vector in z-direction), while the in-plane components are equal to zero. This condition is satisfied for the rigid substrate ($E \geq 30\ kPa$). In this case, the force of gravity cannot deform the substrate under the aggregate body. However, the aggregate weight can deform softer substrate. The ability of the cell aggregate to deform the substrate under its body weight has two consequences:

- This substrate deformation induces the generation of in-plane components of the force of gravity which can be unequal. The consequence of produced asymmetric deformation is anisotropic cell spreading within the cell monolayer (i.e. $g_x \neq g_y$).
- The substrate deformation results in an increase in the contact area between the aggregate and substrate, while the force of gravity is the same, and on that base causes a decrease in the compressive stress on the cells at the aggregate-substrate contact area. Consequently, cells at the contact area undergo unjamming which results in an increase in the velocity of aggregate center of mass $\vec{v}_a$, while the cell aggregate is immobile on the rigid substrate.

The mixing force is formulated here as a function of the gradient of the Gibbs free energy of interactions $\vec{F}_m^{c-M} = \frac{1}{\Delta_c} \vec{\nabla} g_{Mc}{}^c$ (where $\Delta_c$ is the thickness of the perturbed layer of the epithelial cells at the contact area between the cell aggregate and substrate which is an order of magnitude larger than the size of single cells). The surface interaction between cells and substrate matrix can be expressed by Gibbs free energy of interactions $g_{e-M}$ per surface which is equal to:

$$g_{c-M} = g_{M-c}^{c} + g_{M-c}^{M} \tag{8}$$

where $g_{M-c}^{c}$ is the contribution of cells equal to $g_{M-c}^{c} = \gamma_{M-c} - \gamma_{c}$ and $g_{M-c}^{M}$ is the contribution of matrix equal to $g_{M-c}^{M} = \gamma_{M-c} - \gamma_{M}$, $\gamma_{M}$ is the surface tension of matrix, $\gamma_{c}$ is the surface tension of cells, and $\gamma_{M-c}$ is the interfacial tension between cell aggregate and the matrix. The mixing force can stimulate the movement of cells: (1) from the aggregate bulk region toward the cell-substrate contact area (i.e wetting) for $\vec{F}_{m}^{c-M} > 0$, or (2) from the contact area toward the aggregate bulk region for $\vec{F}_{m}^{c-M} < 0$ (i.e. de-wetting). Therefore interfacial tension force accompanied by the mixing force and the force of gravity also drives cell wetting.

The interfacial tension force is expressed as: $n_{c}\vec{F}_{it}^{c-M} = n_{c}S^{c-M}\vec{u}_{c}$ (where $\vec{u}_{c}$ is the displacement field caused by movement of cells, $S^{c-M}$ is the cell spreading coefficient. If the $S^{c-M} > 0$, the interfacial tension force $n_{c}\vec{F}_{it}^{c-M}$ drives cell wetting, while for $S^{e-M} < 0$ this force drives the de-wetting. Cells firstly undergo extension (wetting) with the corresponding spreading coefficient $S^{c-M} = \gamma_{M} - (\gamma_{c} + \gamma_{Mc}) > 0$, which means that $\gamma_{M} > \gamma_{Mc}$. The epithelial wetting causes stretching of AJs which leads to their reinforcement (Devanny et al., 2021). The reinforcement of AJs results in an increase in the surface tension $\gamma_{c}$. When the tissue surface tension $\gamma_{c}$ becomes high enough to ensure that $\gamma_{M} < \gamma_{c} + \gamma_{Mc}$ and $S^{c-M} < 0$, then cell aggregate undergoes compression (i.e. de-wetting). The de-wetting causes an increase in cell packing density $n_{c}$ and consequently intensifies CIL which leads to a decrease in the surface tension $\gamma_{c}$ (Pajic-Lijakovic and Milivojevic, 2022b). Consequently, the Gibbs free energy of interactions of epithelial cells $g_{M-c}^{c}$ is $g_{M-c}^{c} > 0$ for the wetting and $g_{M-c}^{c} < 0$ for the de-wetting.

The viscoelastic force represents the consequence of an inhomogeneous distribution of cell residual stress at the aggregate-substrate contact area and within the cell monolayer and has been formulated by Murray et al. (1988): $\vec{F}_{Tvc} = \vec{\nabla}(\tilde{\sigma}_{cr} - \tilde{\sigma}_{Mr})$ (where $\tilde{\sigma}_{cr}$ is the cell residual stress accumulated within the cell monolayer equal to $\tilde{\sigma}_{cr} = \tilde{\sigma}_{crV} + \tilde{\sigma}_{crS}$, $\tilde{\sigma}_{crV}$ and expressed by eq. 2, $\tilde{\sigma}_{crS}$ is the cell shear residual stress expressed by eq. 6, $\tilde{\sigma}_{Mr}$ is the matrix residual stress accumulated within the cell monolayer equal to $\tilde{\sigma}_{Mr} = \tilde{\sigma}_{MrV} + \tilde{\sigma}_{MrS}$, $\tilde{\sigma}_{MrV}$ is the matrix normal residual stress expressed by eq. 5, $\tilde{\sigma}_{MrS}$ is the matrix shear residual stress expressed by eq. 7). The viscoelastic force is directed always opposite of epithelial cell movement in order to reduce it (Pajic-Lijakovic and Milivojevic, 2020). The reduction of epithelial cell movement is caused by the cell residual stress accumulation. Tse et al. (2012) revealed that compressive stress of ~774 $Pa$ reduced the movement of MCF-10A cells. Riehl et al. (2020) pointed out that shear stress of 1.5 Pa reduces movement of MCF-10A epithelial cells. However, this stress stimulates movement of mesenchymal cancerous breast cells. Accumulation of cell normal residual stress induces an increase in cell packing density which can lead to the cell jamming (Trepat et al., 2009). Cell movement reduction induces a decrease in the viscoelastic force which intensifies cell movement again (Pajic-Lijakovic and Milivojevic, 2022a). Consequently, the viscoelastic force is responsive for the oscillatory trend of cell wetting and de-wetting.

Beside the viscoelastic force, the traction force $\rho_{c-M}\vec{F}_{tr}^{c-M}$ acts to reduce cell movement on the matrix. This reduction is pronounced for higher density and the strength of FAs [50]. Murray et al. [51] formulated the traction force as: $\rho_{c-M}\vec{F}_{tr}^{c-M} = \rho_{c-M}k_c\vec{u}_M^{\ c}$ (where $\rho_{c-M}$ is the density of FAs, $k_c$ is the elastic constant per single FA, and $\vec{u}_M^{\ c}$ is the matrix displacement field caused by cell traction). Cells are able to establish FAs on the rigid substrate and the substrate of the mediated rigidity (Beauene et al., 2018). However, cells cannot establish FAs on the soft substrates (i.e. $E < 7\ kPa$) and undergo random single cell movement (Beauene et al., 2018). Under this condition, the tissue surface tension $\gamma_c \to 0$ and cells cannot undergo de-wetting (Beauene et al., 2018).

The force balance of cell wetting on the rigid substrate (and the substrate of moderated rigidity) can be expressed by modifying the model proposed by Pajic-Lijakovic and Milivojevic (2020;2022e). In order to describe the cell wetting, we introduced here the force of gravity, mixing force and modified the surface tension force in the form of the interfacial tension force. The corresponding force balance is expressed as:

$$\langle m \rangle_c n_c(\Re,\tau)\frac{D\vec{v}_c(\Re,\tau)}{D\tau} = \vec{F}_{ga} + \vec{F}_m^{c-M} + n_c\vec{F}_{it}^{c-M} - \vec{F}_{Tve}^{c-M} - \rho_{c-M}\vec{F}_{tr}^{c-M} \qquad (9)$$

where $\Re = \Re(x,y,z)$ represents the coordinate of cells within the aggregate-substrate contact area and within the cell monolayer, $\langle m \rangle_c$ is the average mass of single cells, $n_c$ is the packing density of cells, the cell velocity is equal to $\vec{v}_c(\Re,\tau) = \vec{v}_a + \vec{v}_F(\Re,\tau)$, $\vec{v}_a$ is the velocity of the aggregate center of mass, $\vec{v}_F(\Re,\tau)$ is the velocity of cells within the monolayer, and $\frac{D\vec{v}_c}{D\tau} = \frac{\partial \vec{v}_c}{\partial \tau} + (\vec{v}_c \cdot \vec{\nabla})\vec{v}_c$ is the material derivative (Bird et al., 1960). The velocity $\vec{v}_a \approx 0$ for the rigid substrate, while at softer substrate, the velocity increases.

While the force of gravity drives cell wetting, this force reduces cell de-wetting. The interfacial tension force changes the direction for the case of the de-wetting in comparison to cell wetting. The mixing force also changes the direction and drives the cell de-wetting accompanied by the interfacial tension force. The viscoelastic force is capable of reducing every cell movement caused by an increase in the cell residual stress. The corresponding force balance for cell de-wetting on the substrate can be expressed by modifying the model proposed by Pajic-Lijakovic and Milivojevic (2020;2022e) as:

$$\langle m \rangle_c n_c(\Re,\tau)\frac{D\vec{v}_c(\Re,\tau)}{D\tau} = \vec{F}_m^{c-M} + n_c\vec{F}_{it}^{c-M} - \vec{F}_{ga} - \vec{F}_{Tve}^{c-M} - \rho_{c-M}\vec{F}_{tr}^{c-M} \qquad (10)$$

Consequently, the viscoelastic force is responsible to cell oscillatory wetting (extension) and oscillatory de-wetting (compression), while the tissue surface tension is the key parameter which induces wetting/de-wetting (Douezan et al., 2011). The competition between forces which stimulate cell movement and the forces which reduce the movement induces oscillatory change of velocity for both sub-populations. The scenario of cell rearrangement within the cell aggregate-substrate contact area as well as within the cellular monolayer can be summarized as follow:

- Cells undergo movement from the aggregate bulk toward the aggregate-substrate contact area driven by the force of gravity, mixing force and interfacial tension force. Resulted cell extension induces cell residual stress accumulation which leads to an increase in the viscoelastic force.
- The viscoelastic force reduces movement of epithelial cells. A decrease in cell velocity results in the relaxation of cellular system which leads to a decrease in the viscoelastic force. Then cells start movement again in the same direction if and only if following relations exist $\gamma_c < \gamma_{cM}$ and $\gamma_M > (\gamma_c + \gamma_{Mc})$.
- Cell extension induces an increase in the tissue surface tension $\gamma_c$. If the epithelial surface tension $\gamma_c$ becomes high enough to satisfy the relations $\gamma_c > \gamma_{cM}$ and $\gamma_M < (\gamma_c + \gamma_{Mc})$, cells undergo de-wetting.
- Cell compression during the de-wetting induces the accumulation of compressive cell residual stress which leads to an increase the viscoelastic force capable of reducing cell movement. Cell movement reduction results in a decrease in the viscoelastic force and the cell movement can start again in the same direction.
- The compression intensifies cell-cell interactions, as well as, the CIL which causes a weakening of AJs accompanied by a decrease in the tissue surface tension. When the tissue surface tension becomes lower than the interfacial tension i.e. $\gamma_c < \gamma_{cM}$ and $\gamma_M > (\gamma_c + \gamma_{Mc})$, the extension can start again.

Besides the cell force balance, it is necessary to formulate the cell mass balance in order to describe corresponding cell rearrangement.

### 3.6 The mass balance

The mass balance of epithelial cells during the wetting and de-wetting can be expressed by modified the model proposed by Murray et al. (1988) for free expansion of cell monolayers. We introduced here the durotaxis flux (Pajic-Lijakovic and Milivojevic, 2022c) and the Marangoni flux (Pajic-Lijakovic and Milivojevic, 2022d). The mass balance can be expressed as:

$$\frac{\partial n_c(\Re,\tau)}{\partial \tau} = -\vec{\nabla} \cdot \left(\vec{J}_{conv} + \vec{J}_{cond} + \vec{J}_d + \vec{J}_{Mc}\right) \qquad (11)$$

where the convective flux is equal to $\vec{J}_{conv} = n_c \vec{v}_c$ (for the convective regime) and the conductive flux is equal to $\vec{J}_{cond} = -D_{eff} \vec{\nabla} n_c$ (for the conductive regime), $D_{eff}$ is the effective diffusion coefficient. The durotaxis flux which describes the directed CCM, caused by the matrix stiffness gradient, can be expressed as $\vec{J}_d = k_d n_c \Delta V_m (\vec{\nabla} E_m + \vec{\nabla} G_m)$ (where $k_d$ is the model parameter which represents a measure of matrix mobility induced by cell action, $\Delta V_m$ is the volume of a matrix part, $E_m$ is the matrix Young's modulus, and $G_m$ is the matrix shear modulus) (Pajic-Lijakovic and Milivojevic, 2022c). The matrix stiffness gradient can be generated during CCM depending on the matrix viscoelasticity (Pajic-Lijakovic et al., 2022). The Marangoni flux $\vec{J}_{Mc}$ is equal to $\vec{J}_{Mc} = k_{Mc} n_c \vec{\nabla} \gamma_{M-e}$, $k_{Mc}$ is the model parameter which quantifies the mobility of epithelial cells (Pajic-Lijakovic and Milivojevic, 2022d). The Marangoni flux describes the movement of epithelial cells from the aggregate bulk region toward the aggregate-substrate contact area (i.e. wetting) as the consequence of the established interfacial tension gradient such that $\gamma_c > \gamma_{cM}$. An increase in the epithelial surface tension caused by the monolayer extension leads to a change in the direction of the interfacial tension gradient, for the case when $\gamma_c < \gamma_{cM}$, which stimulates cell movement from the aggregate bulk toward the cell-matrix contact area (i.e. de-wetting).

Cell spreading within a monolayer around the cell aggregate on rigid substrate corresponds to the convective regime for the cell packing density $n_c \leq n_{conf}$ and cell velocity in the range $0.1 \frac{\mu m}{min} < \vec{v}_c < 1 \frac{\mu m}{min}$ based on the experimental findings by Beaune et al. (2018) and Pérez-González et al. (2019). However, the local cell residual stress accumulation caused by CCM can induce local change from convective to conductive (diffusion) mechanism of cell movement and even cell jamming transition (Pajic-Lijakovic and Milivojevic, 2021).

In further consideration, we would like to discuss oscillatory dynamics of cell wetting/de-wetting in the context of low Reynolds turbulence by extracting the relevant dimensionless criteria.

### 4. Low Reynolds number turbulence induced by cell aggregate wetting

Appearance of the turbulence has been related to an inhomogeneous transfer of energy caused by flow of liquids at high Reynolds number (i.e. $R_e \sim 10^5$). The $R_e$ number is equal to $R_e = \frac{vL\rho}{\eta}$ (where $v$ is the velocity, $\eta$ is the viscosity, $L$ is the characteristic length, and $\rho$ is the density). Inhomogeneous energy transfer causes the generation of instabilities in the form of swirls which lead to an oscillatory change of liquid velocity. These inertial effects represent the main characteristic of the turbulence. However, the turbulence also can be generated during flow of various soft matter systems such as flexible, long-chain polymer solutions at low $R_e$ number (Groisman and Stainberg, 1998;2000). In this case, the turbulence is induced by the system viscoelasticity. It is so called "elastic turbulence". The viscoelasticity is related to the stress relaxation, strain change under flow, and residual stress accumulation. These processes can occur at the same time scale or at various time scales. Groisman and Steinberg (1998;2000) introduced the Weissenberg number $W_i$ to characterize the elastic turbulence caused by stretching of polymer chains under flow. The $W_i$ number correlates viscoelasticity with the degree of anisotropy caused by the rearrangement of the system constituents. Groisman and Steinberg (1998) expressed the $W_i$ number in the form $W_i = \tau_R \dot{\varepsilon}$ (where $\tau_R$ is the stress relaxation time and $\dot{\varepsilon}$ is the rate of deformation). The stress relaxation time corresponds to the characteristic time for the rearrangement of polymer chains and accounts for the system viscoelasticity in this case. The $W_i$ number is $\sim 1$ which suggests that the stress relaxation time and the deformation rate are the same order of magnitude (Groisman and Stainberg, 1998;2000).

The low $R_e$ number turbulence also occurs during the rearrangement of multicellular systems caused by CCM (Serra-Picamal et al., 2012; Notbohm et al., 2016; Pajic-Lijakovic and Milivojevic, 2020;2022b). In contrast to other soft matter systems, the multicellular systems are active, capable of self-rearranging, which has been treated as an "active turbulence" (Alert et al., 2021). The main characteristic of the viscoelasticity of multicellular systems caused by CCM is their multi-time nature (Marmottant et al., 2009; Pajic-Lijakovic and Milivojevic, 2020). Cell movement, strain caused by this movement, and the residual stress accumulation occur at time scale of hours, while the stress relaxation time corresponds to a time scale of minutes (Marmottant et al., 2009; Pajic-Lijakovic and Milivojevic, 2019a;2020). Therefore corresponding $W_i$ number is $W_i \ll 1$ (Pajic-Lijakovic and Milivojevic, 2022b). The effective Weissenberg number can be formulated in the case of cellular systems as $W_{i\ eff} = t_p \dot{\varepsilon}$ (where $t_p$ is the cell persistence time which corresponds to the time scale of several tens of minutes to hours (Mc Cann et al., 2010)). The time $t_p$ could be related to the degree of anisotropy or orientation generated by the deformation which is caused by CCM rather than the stress relaxation time. The cell persistence time depends on the cell residual stress and matrix residual stress accumulation. Pérez-González et al. (2019) revealed that trend of cell movement ordering, which results in lower cell residual stress accumulation, represents the characteristic of lower sized cell aggregate. The matrix residual stress accumulation results in the matrix stiffening, which stimulates the cell movement ordering and could be quantified by higher value of the cell persistent time.

In order to characterized oscillatory trend in cell aggregate wetting/de-wetting on rigid substrates, additional dimensionless number is needed besides $R_e$ number and $W_{i\ eff}$ number. It could be the Weber number equal to $W_e = \frac{v^2 L \rho}{\gamma_c}$. This number represents a measure of inertia effects relative to the surface tension.

## 5. Conclusions

Over the last years it has become clear that epithelial spreading during morphogenesis and tissue regeneration involve oscillatory changes in tissue morphology. These oscillatory changes of the geometry of multicellular systems, cell velocity, resulted strain, and residual stress accumulation have been interpreted in the form of mechanical waves. The mechanical waves are related to effective long time inertial effects and represent an integral part of 2D and 3D CCM and have been treated as the low Reynolds number turbulence. The deeper insight into the mean features of physical interactions between epithelium and substrate matrix is essential for the establishment of key physical parameters responsible for the appearance of the low Reynolds number turbulence. This oscillatory dynamics of cell rearrangement is considered on the model system such as wetting/de-wetting of cell aggregate on rigid substrates which includes cell aggregate movement and isotropic/anisotropic spreading of cell monolayer around the aggregate depending on the substrate rigidity and aggregate size. This model system accounts for transition between 3D epithelial aggregate and 2D cell monolayer as a product of: (1) tissue surface tension, (2) surface tension of substrate matrix, (3) cell-matrix interfacial tension, (4) interfacial tension gradient, (5) viscoelasticity caused by CCM, and (6) viscoelasticity of substrate matrix. These physical parameters vary depending on the cell contractility and state of cell-cell and cell matrix adhesion contacts, as well as, the stretching/compression of cellular systems caused by CCM. The main results discussed in this review were obtained by combining constitutive models with biological and bio-mechanical experiments, and we can summarize them as follows:

- The key control parameters responsible for oscillatory trend in cell aggregate wetting/de-wetting are the epithelial surface tension, cell residual stress, and matrix residual stress. The epithelial surface tension depends on the state of AJs between contractile (migrating) cells. While the tissue surface tension change causes cell wetting and de-wetting, the cell residual stress accumulation is responsible for oscillatory changes of both processes, the wetting and de-wetting.
- The wetting of cell aggregate induces stretching of AJs which leads to an increase in the tissue surface tension. When the tissue surface tension becomes high enough to ensure that cohesion energy among cells is higher than cell-matrix adhesion energy, the cell de-wetting starts to appear. The compression caused by cell de-wetting intensifies CIL which results in a decrease in the epithelial surface tension. Then the wetting can start again.
- Cell wetting/de-wetting results in the cell residual stress accumulation. This stress accumulation depends on: cell-matrix interfacial tension, interfacial tension gradient, viscoelasticity caused by CCM, and the cell aggregate gravitation. The accumulation of the residual stress within the epithelium is capable of reducing cell movement which leads to a decrease in the cell residual stress. Then cell movement starts again. The cell residual stress accumulation is pronounced for larger aggregates and cell spreading on the more rigid substrates.
- Matrix residual stress accumulation caused by cell tractions and gravitation of cell aggregate represents a product of the matrix viscoelasticity. This stress accumulation results in the matrix stiffening capable of intensifying cell spreading.
- Oscillatory trend of cell wetting/de-wetting was discussed in the context of low-Reynolds turbulence. Besides the Reynolds number, effective Weissenberg and Weber numbers are proposed for characterizing this type of turbulence.

Additional experiments are needed in order to correlate the cell-matrix interfacial tension and the interfacial tension gradient with the cell residual stress accumulation and to measure the epithelial surface tension vs. time during the cell aggregate wetting/de-wetting.

**Acknowledgements**: This work was supported by the Ministry of Education, Science and Technological Development of the Republic of Serbia (Contract No.451-03-68/2022-14/200135).

**Declaration of interest**: The author reports no conflict of interest.